\documentstyle[sprocl]{article} 

\def\Journal#1#2#3#4{{#1} {\bf #2}, #3 (#4)} 
 
\def\NPB{{\em Nucl. Phys.} B} 
\def\PLB{{\em Phys. Lett.}  B} 
 
\def\PRD{{\em Phys. Rev.} D} 
\def\ZPC{{\em Z. Phys.} C} 

\def\be{\begin{equation}} 
\def\ee{\end{equation}} 
\def\bea{\begin{eqnarray}} 
\def\eea{\end{eqnarray}} 
\def\ha{{1\over 2}}
\def\normord#1{\mathopen{\hbox{\bf:}}#1\mathclose{\hbox{\bf:}}} 
\def\bra#1{\lan#1|}
\def\ket#1{|#1\ran}
\def\b{\beta}
\def\d{\delta}
\def\frac#1,#2{{#1\over #2}}
\def\lan{\langle}
\def\ran{\rangle}
\def\fr#1,#2{{#1\over #2}}
 
\begin{document} 
 
\title{SUBTLETIES IN THE LIGHT-CONE REPRESENTATION} 
 
\author{GARY MCCARTOR} 
 
\address{Department of Physics, SMU, Dallas, Texas, USA} 
 
 
\maketitle\abstracts{ To produce an isomorphism between the light-cone
and equal-time representations some additional formalism beyond that
originally proposed for the light-cone representation may sometimes be
required.  The additional formalism usually involves zero modes and is
most likely to affect delicate, high energy aspects of the solution
such as condensates.  In this talk I will review some of the
information which has been obtained in the past few years on these
issues with particular emphasis on the Schwinger model as an example.}
 
\section{The Light-Cone Representation} 
 
To form the light-cone representation for a covariant field theory one
proceeds as follows: One specifies initial conditions ( the canonical
commutation relations ) on a characteristic surface, typically $x^+
\equiv x^0 + x^1 = 0$; Negative frequency Fourier modes taken along
this surface are taken to be creation operators while positive
frequency modes are destruction operators; The bare vacuum is the
state destroyed by all the positive frequency modes. The dynamical
operators are calculated by integrating densities over the initial
value surface, for instance: 
$$
      P^- = \ha\int\normord{T^{-+}}dx^- d^2x^\perp
$$

The relation of this representation to the equal-time representation (
initial conditions on the surface $x^0 = 0$ ) is rather simple for
free theories but extremely complicated for interacting theories.
Particularly in view of what I shall say below, it is appropriate to
provide some reasons for being interested in the light-cone
representation.  Some commonly given reasons are:

\hskip1truein{$\bullet$} Boosts are simple ( theorem )

\hskip1truein{$\bullet$} The vacuum is simple ( ``theorem'' )

\hskip1truein{$\bullet$} Closer to partons ( hope )

The first bullet is a theorem; boosts are translations within the
initial value surface and thus kinematical and thus simple.  The
second bullet, the physical vacuum is the bare vacuum, is sometimes
given as a theorem but it is not; in particular, as we shall see
below, it isn't true.  The third bullet lists the expectation that the
light-cone bare states are closer to the partons that are observed to
make up hadrons than the equal-time bare states. While arguments, some
of them quite sensible, are given to defend this expectation, at the
moment it is nothing more than a hope.

At the same time that there have been these hopes for advantages to
the light-cone representation there have been a number of issues,
perhaps one should say puzzles, which have cast doubt as to whether
the light-cone representation space could serve the same physics as
the equal-time representation space.  Some of these puzzles are:

\hskip1truein{$\bullet$} Degenerate vacua ?

\hskip1truein{$\bullet$} Condensates ?

\hskip1truein{$\bullet$} Causality ?

The first bullet is to be seen in conjunction with the second bullet
of the earlier list.  If there are more than one possible ground
states, and sometimes there are, the bare vacuum could be one of the
possibilities but not all.  A somewhat less concrete puzzle is
presented by condensates: in the equal-time representation it is usual
to think of them as vacuum phenomena.  If they do not appear in that
way in the light-cone representation, how do they appear?  Finally,
What about the general question of causality --- since the points on
our initial value surface are causally connected, can we really
specify initial conditions on them without great care to avoid
contradictions?

\section{Zero Modes}

Perhaps the greatest interest in the light-cone representation stems
from the idea that the physical vacuum is the bare vacuum.  At the
same time this idea leads to the most direct contradictions as
discussed in the previous section.  Let us therefore review the
argument that $\ket{\Omega} = \ket{0}$.  The argument begins with the
statement that the operator $P^+$ in the interacting theory is the
same as in free theory.  The physical vacuum should satisfy $
P^+\ket{\Omega} = 0$ and, at least for theories which can be
initialized on $x^+ = 0$, there is only one such state --- the bare
vacuum.  That's the essence of the argument.  The argument that $P^+ =
P^+_{FREE}$ may be given two different ways.  The first way just
calculates it by integrating a density over the initial value surface:
$$
              P^+ =  \ha \int T^{++} dx^-
$$
$$
        T^{++} = \sum_{\phi} \fr{\partial \phi},{\partial x_{+}}
\fr{\partial {\cal L}},{\partial (\partial_{+} \phi)} 
- g^{++} {\cal L}          
$$
Since $g^{++} = 0$ we get $P^+ = P^+_{FREE}$.  More pertinent to the
remarks below is the following argument: The fields must satisfy the
Heisenberg relation
$$
                   \partial_-\phi = {i\over 2}[P^+,\phi]
$$
This relation may be checked using values entirely within the initial
value surface.  But since we initialize the fields to be isomorphic to
free fields on that surface, we know that there
$$
                   \partial_-\phi = {i\over 2}[P^+_{FREE},\phi]   
$$
Thus if the fields form an irreducible set, which basically means we
have a well posed physics problem, the only modification to
$P^+_{FREE}$ we can make would be to add a multiple of the identity.
But, of course, that is exactly the point: for many cases one cannot
completely specify the physics problem by providing information on the
characteristic $x^+ = 0$~\cite{m1}.  If we consider the free, massless
Fermi field in two dimensions with periodicity conditions to induce a
discrete set of modes we find
$$
    \psi_+(x^+,x^-) = {1\over\sqrt{2L}}\sum_{n=1}^\infty 
   b(n) e^{-ik_-(n)x^-} +
   d^*(n) e^{ik_-(n)x^-}       
$$
$$
    \psi_-(x^+,x^-) = {1\over\sqrt{2L}}\sum_{n=1}^\infty 
   \beta (n) e^{-ik_+(n)x^+} +
   \delta ^*(n) e^{ik_+(n)x^+}      
$$
Here we see that the entire field, $\psi_-$, is composed of zero modes
--- functions only of $x^+$.  The $\psi_-$ field cannot be initialized
on the surface $x^+ = 0$.  Note also that the formula for $P^-$ given
above will not suffice; we need
$$
      P^- = \ha\int\normord{T^{-+}}dx^- + \ha\int\normord{T^{--}}dx^+
$$

The $\b$ and $\d$ modes are examples of unconstrained zero modes ---
that is, they are true degrees of freedom of the system.  It is worth
remarking that there is another type of zero mode called a constrained
zero mode.  These are not true degrees of freedom but are related to
the degrees of freedom by constraint relations.  An example is
furnished by $\phi^4$ theory.  In two dimensions the equation of
motion is 
$$
         (4\partial_+\partial_- + \mu_0^2)\phi = -\lambda\phi^3
$$
Even though the $\phi$ field does not have a zero mode in free theory,
the right hand side of the above equation does have one; thus in the
interacting theory the field $\phi$ does have a zero mode which goes
to zero when $\lambda$ goes to zero and which is determined in terms
of the degrees of freedom by the equation of motion.  This type of
zero mode was first discussed by Maskawa and Yamawaki~\cite{my}.  Such
a zero mode is responsible for the condensate in wrong-sign $\phi^4$
theory~\cite{dr}~\cite{ew}.  Thus we see that in fact aspects of the
solution which appear as vacuum structure in the equal-time
representation can appear in other ways in the light-cone
representation --- sometimes as the development of a constrained zero
mode.  The constrained zero modes cannot lead to the existence of
degenerate ground states, however.  For that we shall require the
unconstrained zero modes such as those of the $\psi_-$ field.

\section{The Schwinger Model}

Let us now illustrate some of these ideas with the best studied case:
the Schwinger model~\cite{m3}.  First consider the problem of
calculating the operator $P^+$. We shall choose antiperiodic boundary
conditions for the Fermi fields and initialize $\psi_+$ on $x^+ = 0$
and $\psi_-$ on $x^- = 0$ 
$$
    \psi_+(0,x^-) = {1\over\sqrt{2L}}\sum_{n=1}^\infty 
   b(n) e^{-ik_-(n)x^-} +
   d^*(n) e^{ik_-(n)x^-}       
$$
$$
    \psi_-(x^+,0) = {1\over\sqrt{2L}}\sum_{n=1}^\infty 
   \beta (n) e^{-ik_+(n)x^+} +
   \delta ^*(n) e^{ik_+(n)x^+}      
$$
The general considerations based on the Heisenberg equations suggest
that modes from the $\psi_-$ field might mix with $P^+$.  An explicit
calculation shows that this expectation is realized.  We find
$$
     P^+ =  \ha \int^L_{-L} \normord{2i \Bigl(\psi^*_+ \partial_- 
\psi_+ - \partial_- \psi^*_+ \psi_+\Bigr)} dx^-
$$
This equation seems to have the same form as free theory but the
symbols mean something different.  To maintain gauge invariance we
must define
$$
T^{++} =  2i \lim_{\epsilon^-\rightarrow0} \Biggl(
 e^{-ie\int_x^{x+\epsilon^-} A_-^{(-)} dx^-}
\psi_+^*(x+\epsilon^-) \partial_- \psi_+(x)
e^{-ie\int_x^{x+\epsilon^-} A_-^{(+)} dx^-} 
 -  {\rm C.C.}-{\rm V.E.V.} \Biggr) 
$$
With this definition a simple calculation gives
$$
      P^+ = P^+_{FREE} - (\ha A^+)^2
$$
The operator $A^+$ is an example of a constrained zero mode.  To
understand how to calculate it properly we shall require some
discussion.  If we take the zero mode component of the Maxwell
equation
$$
	\fr{\partial^2 A^-},{\partial x^{-2}} = -\ha {J}^+ 
$$
we get
$$
        \fr {1},{2L}Q_+ = \fr {e^2},{2\pi}A^+ \qquad ( WRONG )
$$
If this relation were correct we would have an immediate problem in
that $Q_+$ is composed of modes from the $\psi_+$ field which are the
ones we saw above could NOT be mixed with $P^+$ without producing an
immediate conflict with the Heisenberg equations.  To understand why
the relation is wrong we can consider a problem in classical
E\&M~\cite{m2}.  A charge density which is constant everywhere in
space cannot create a field.  Among other things there would be no
preferred direction for the field to point.  On the other hand, if
$\rho(0)$ is a charge density constant in space, the Maxwell equation
$$
	{\partial E \over \partial x}=\rho(0)
$$
does not have zero as a possible solution.  The field really couples
to the current $J^\prime$ where
$$
          {J^\prime}^0 = J^0 - J^0(0)\quad ;\quad  {J^\prime}^1 = J^1
$$
With this correction the solution for $A^+$ becomes
$$
   \fr {1},{2L}Q_- = \fr {e^2},{2\pi}A^+
$$
Which then gives
$$
P^+ =  \ha \int^L_{-L} \normord{2i \Bigl(\psi^*_+ \partial_- \psi_+ 
- \partial_- \psi^*_+
\psi_+\Bigr)} dx^-  = P^+_{FREE} - {1\over 4Lm^2}Q_-^2
$$
Now the operators mixed with $P^+$ come from the $\psi_-$ field and
there is no conflict with the Heisenberg equations.  The correction to
$P^+$ --- necessary to maintain gauge invariance --- causes some
states which are split in free theory to become degenerate ground
states in the interacting theory; linear combinations of them provide
the expected $\theta$ states familiar in the Schwinger model.  The
physical subspace of the Schwinger model is chargeless and so defined
by
$$
        \ha(Q_+ + Q_-)\ket{p} = 0
$$
So we may write
$$
           Q_- \approx Q_+
$$
From this point of view the correction from the equation labeled
$WRONG$ to the correct result can be viewed as substituting an
operator for its weak equivalent.  In the case of the Schwinger model
we completely understand the physical basis for that requirement; in
some other cases we find that we must make such a substitution, again
to avoid conflict with the Heisenberg equations, and we do not as
completely understand the physical basis.  General rules on the
subject are not known and we currently have to treat each case
separately.

A full operator solution can be given with these boundary conditions.
Due to space limitations I shall not do so here but will list some of
the things one must do to formulate such a solution~\cite{m3}

\hskip1truein{$\bullet$} KEEP $\psi_-(x^+ + 2L) = -\psi_-(x^+)$

\hskip1truein{$\bullet$} COUPLE TO $J^\prime$

\hskip1truein{$\bullet$} KEEP $A^+$ --- A CONSTRAINED ZERO MODE

\hskip1truein{$\bullet$} MAINTAIN GAUGE INVARIANCE

\hskip1truein{$\bullet$} PHYSICAL SUBSPACE $D(n)\ket{P} = 0$

\hskip1truein{$\bullet$} CALCULATE $P^-$ USING $x^+\ and\ x^-$

The next to last bullet refers to the fact that a more stringent
physical subspace condition must be imposed in this gauge than just
the chargeless one.  Some properties of the solution one then finds
include~\cite{m3}

\hskip1truein{$\bullet$} SPECTRUM

\hskip1truein{$\bullet$} ANOMALY

\hskip1truein{$\bullet$} $\theta$--VACUUM STRUCTURE

\hskip1truein{$\bullet$} MUCH SIMPLER SOLUTION ($a^*(n) = C^*(n))$)

\hskip1truein{$\bullet$} $\bra{\Omega} \bar{\psi} \psi \ket{\Omega} =
\fr{1},{L} cos\theta$

\looseness=-1\noindent The first three bullets are in accord with
standard calculations.  The fourth bullet makes definite the idea that
the light-cone bare states are closer to partons: the light-cone
fusion operators create the Schwinger particles.  The last bullet
shows an unexpected result: the condensate goes to zero as $L$ goes to
infinity.  That aspect is not due either to the choice of light-cone
gauge or to the choice of quantization surface but rather to the
choice of boundary conditions: the continuum the light-cone gauge
solution has the expected value of the condensate whatever
quantization surface is used.  The imposition of the periodicity
conditions has too abruptly removed the small $p^+$ region of the
spectrum and the very singular behavior there, which is responsible
for the condensate, in not recovered as $L$ becomes large.

I believe that the behavior of the condensate we have just seen may be
related to a puzzle in the 't Hooft~\cite{th} model (Large N
$QCD_{1+1}$).  't Hooft solved the model and gave the ( dressed )
propagator.  By Fourier transforming the propagator to coordinate
space one may easily determine that it implies no condensate.  On the
other hand Zhitnitsky~\cite{az} has shown that the spectrum and
wavefunctions 't Hooft derived from his propagator imply that $$
\bra{\Omega} \bar{\psi} \psi \ket{\Omega} = -N
\sqrt{\fr{g^2N},{12\pi}} $$ 't Hooft used light-cone gauge and
light-cone quantization to formulate his solution.  While he did not
use periodicity conditions, his method of regulating the small $p^+$
region was pretty abrupt and had much the same effect.  The situation
is further complicated by the fact that T.T. Wu~\cite{tw} has given a
different propagator, the difference being solely due to the treatment
of the small $p^+$ region.  Some recent discussion of those issues is
in Bassetto and Griguolo~\cite{ab}.

\section{Discussion}

In view of the difficulties I have presented one might well ask: ``can
all this really be worth it?''  I do not yet know; but

\hskip1truein{$\bullet$} Boosts are simple

\noindent In cases where the answer is known

\hskip1truein{$\bullet$} The vacuum is simpler

\noindent Not completely trivial but much simpler.  And at least for
the Schwinger model it is true that the light-cone bare states are

\hskip1truein{$\bullet$} Closer to partons

\end{document}